\documentclass[a4paper,reqno,12pt,draft]{article}
\usepackage{amssymb,euscript,bbold}
\newcommand{\be}{\begin{equation}}
\newcommand{\ee}{\end{equation}}
\newcommand{\ba}{\begin{eqnarray}}
\newcommand{\ea}{\end{eqnarray}}

\newcommand{\ft}{\footnote}

\begin{document}
\input{epsf}

\begin{flushright}
RUNHETC-2003-24\\
CU-TP-1092
%26th May 1998
\end{flushright}
\begin{flushright}
%{\sf \today}
\end{flushright}
\begin{center}
\Large{\sc Freund-Rubin Revisited}\\
\bigskip
\bigskip
\large{\sc B.S. Acharya$^{a,b}$, F. Denef$^{a,b}$, C. Hofman$^b$, N. Lambert$^b$}
{\renewcommand{\thefootnote}{}
\footnotetext{bacharya, denef, hofman, nlambert@physics.rutgers.edu}}\\
\bigskip\large
{\sf $^a$Department of Physics,\\
Columbia University,\\ 538 W. 120th St.,\\ New York, NY 10027.}
\smallskip\large

{\sf $^b$NHETC, Department of Physics,\\
Rutgers University,\\ 126 Freylinghuysen Road,\\ Piscataway, NJ 08854-0849.}

\end{center}
%\renewcommand{\abstractname}{\sc Abstract}
%\begin{Abstract}
\bigskip
\begin{center}
{\bf {\sc Abstract}}
\end{center}
%\bigskip
%\normalsize
We utilise the duality between $M$ theory and Type IIA string theory
to show the existence of Freund-Rubin compactifications of $M$ theory
on 7-manifolds with singularities supporting chiral fermions.
This leads to a concrete way to study phenomenologically interesting quantum
gravity vacua using a holographically dual three dimensional field theory.
%\end{center}
%\end{abstract}

\newpage

\Large
\noindent
{\bf {\sf 1. Introduction}}
\normalsize
\bigskip

In 1980, Freund and Rubin found a solution of d=11 supergravity which
describes a compactification of the theory on a round seven sphere to
four dimensional anti de Sitter space ($AdS_4$) \cite{fr}. 
We now know that this 
is a vacuum of $M$ theory which arises as the dual to
the three dimensional conformal field theory of $M$2-branes in flat space
\cite{mald}. More generally the Freund-Rubin ansatz is satisfied 
for any seven manifold $X$ whose
metric is Einstein with positive scalar curvature, and these compactifications
are dual to more complicated 3d CFT's \cite{cones}.

Freund-Rubin vacua were originally of interest in a program to obtain
realistic four dimensional physics by appropriately choosing $X$, see
\cite{kk} for a review.
This program originally failed due to Witten's no go theorem which states that
chiral fermions do not arise when $X$ is smooth
\cite{wit}. Moreover, the subsequent
discovery \cite{chsw} that heterotic string compactifications do produce chiral
fermions and a classically zero cosmological constant diverted attention away
from d=11 supergravity compactifications.

More recently, we have learned in the context of compactifications of
$M$ theory on $G_2$-holonomy manifolds to flat space, 
that singularities of $X$ can
support non-Abelian gauge fields \cite{bsa1, bsa2, amv} and
chiral fermions \cite{aw, bw}. 
These are membranes which have collapsed at the
singularity. Many of the arguments which were used to show
the existence of chiral fermions in $G_2$-holonomy compactifications
were local to the singularity, and do not really depend upon $X$ having a
$G_2$-holonomy metric.
Therefore, it is natural to ask if Freund-Rubin compactifications 
with singularities can also produce chiral fermions. 

In this paper we will use the duality between $M$ theory and Type IIA
string theory to show that Freund-Rubin vacua with chiral fermions do indeed
exist. 

Aside from producing new $M$ theory
compactifications with potentially realistic particle
physics, there are several other reasons why this result might be of interest.
As mentioned above, Freund-Rubin  
compactifications are dual to three dimensional
conformal field theories. This offers the attractive possibility of describing,
say, a supergravity GUT using 3d CFT. 
Supersymmetry breaking in four dimensions could also be studied holographically.
Another reason is that, as far as we can 
tell, Freund-Rubin vacua are not dual to compactifications on special 
holonomy manifolds such as the heterotic string on Calabi-Yau threefolds or 
$M$ theory on $G_2$-manifolds.
As such, they might be totally disconnected from these more intensively studied
vacua and thus represent new points on the space of potentially realistic
quantum gravity vacua. This clearly also has implications for the ideas proposed
in \cite{mike}, which presupposes that the special holonomy vacua represent
an `order one fraction' of `realistic vacua'.

Of course, the vacua under discussion here are not realistic since they
have a negative cosmological constant and unbroken supersymmetry.
But the same is true of 
generic ${\cal N}$ $=1$ string or 
$M$ theory compactifications on special holonomy
manifolds  after accounting for appropriate
fluxes and/or quantum corrections, for example see \cite{blah}.
So, in most supersymmetric string or $M$ theory 
compactifications we still need a mechanism for
raising the cosmological constant to, or above, zero. 

In the next section we describe some general properties of
Type IIA duals of Freund-Rubin vacua. We then go on to describe 
a fairly general construction of such Type IIA backgrounds which
are obtained by adding $D$6 and $D$2-branes to $G_2$-holonomy spacetimes.
In section four we describe some examples of such Type IIA
compactifications to $AdS_4$ which contain chiral fermions.
We conclude with a discussion of the holographically  dual description
of these $M$ theory compactifications.

%\newpage
\bigskip
\Large
\noindent
{\bf {\sf 2. IIA Duals of Freund-Rubin Vacua}}
\normalsize
\bigskip

Freund-Rubin compactifications of $M$ theory, on a 7-manifold $X$, arise as the
large $N$ limit of $N$ $M$2-branes residing at the tip of a cone with base $X$.
Specifically, the metric for $N$ $M$2-branes on a cone over $X$ is given by
\be
g_{10+1} = H^{-{2 \over 3}}(r) g_{2+1} + H^{1 \over 3}(r) (dr^2 + r^2 g_7 (X) )
\ee
where $H$ $= 1 + {a^6 \over r^6}$ and $g_7 (X)$ is Einstein with 
cosmological constant equal to 6, $g_{2+1}$ is the d=3 
Minkowski metric and $a^6 \sim \mbox{Vol}(X) N $. 
The solution has non-zero $G$-flux $G = \mbox{dVol}_{2+1}\wedge dH^{-1}$.
Supersymmetry requires that the cone $g_8(C(X)) = dr^2 + r^2 g_7 (X)$ 
is an 8-metric whose holonomy is a subgroup of $Spin(7)$ \cite{cones}. 

In the large $N$ (equivalently small $r$) 
limit the solution becomes the Freund-Rubin compactification
on $X$:
\be
g_{FR} = g(AdS_4 ) + a^2 g_7 (X)
\ee
where $g(AdS_4)$ is the standard metric on $AdS_4$ of ``radius'' $a$.

The $M$2-brane metric (for all $N$) has a dual description in Type IIA 
string theory
if $g_7 (X)$ admits a $U(1)$ action. In this case
\be
g_7 (X) = h^2 (Y) (d\tau + A(Y))^2 + g_6 (Y)
\ee
where $h$ is a function on $Y= X/U(1)$ and $A$ is a 1-form on $Y$.
The dual IIA solution has a metric $g_{IIA}$, dilaton $\phi$ and $RR$
gauge potential $C$ determined from the $M$ theory metric by
\be
g_{10+1} = e^{4\phi \over 3} (d\tau + C)^2 + e^{-{2\phi \over 3}}g_{IIA}
\ee
In particular, the IIA dual of the $M$2-brane metric is
\be
g_{IIA} = r\;h(Y) \Bigl( H(r)^{-{1 \over 2}} g_{2+1} + H(r)^{1 \over 2} dr^2 +
H(r)^{1 \over 2} r^2 g_6 (Y) \Bigr)
\ee
and the dilaton is given by
\be
e^{2\phi \over 3} = r\;h(Y)H(r)^{1 \over 6} 
\ee

At large $r$ the $M$ theory metric becomes the product of 3d Minkowski space
and a Ricci flat cone with base $X$. The large $r$ IIA metric above
is simply the IIA metric one obtains from the large $r$ $M$ theory metric
using the formula $(4)$.

The IIA $D$2-brane metric $(5)$ is similar to the standard $D$2-brane metric
except that in that case there is no overall factor of $rh(Y)$ and 
in the standard case $H(r) \sim 1 + r^{-5}$. The differences are
crucial however, because the large $N$ limit of the $D$2-brane metric $(5)$
is a compactification to $AdS_4$, whereas in the usual case this limit
does not give an $AdS$-background. Specifically, the large $N$ limit 
of $(5)$ gives
\be
{g_{IIA}}|_{N \rightarrow \infty} = a^3h(Y) \Bigl(a^{-6}r^4 g_{2+1}
+ r^{-2}dr^2 + g_6 (Y)\Bigr)
\ee
This is a warped metric on $AdS_4 \times Y$ with a warp factor determined by 
$h(Y)$. Because the large $N$ limit produces an $AdS$ vacuum, the
worldvolume dynamics of the $D$2-branes is conformal in this limit.

Note however that at large $r$ the dilaton is growing, so we cannot
use string perturbation theory to understand the physics. Similarly, at 
small $r$, in the AdS region, $e^{2\phi \over 3} \sim ah(Y)$, so the dilaton
is finite, but
the presence of RR 4-form flux through the $AdS$ makes it difficult
to quantise the theory.

In all the examples we describe in this paper we will actually see that
the strong coupling at large $r$ is an artefact of the failure of
the solution to properly describe the physics there. This will be illustrated
shortly with a simple example.

Another important feature is that when $h$ vanishes along 3-dimensional
submanifolds $Q_i$ of $Y$ the $M$ theory circle goes to zero and
the IIA background has $D$6-branes wrapping $Q_i$. We will mainly
be interested in such examples here.
In these examples, the $M$ theory
circle is contractible unlike, for instance, the case of pure $D$2-branes in
flat space.

As a specific application of these results, which illustrates many features
of the more interesting solutions we will find,
we can consider the standard
$M$2-brane solution, whose large $N$ limit is the original Freund-Rubin 
compactification on the round $S^7$. The metric in $M$ theory takes the
form $(2)$ with
\be
g_7 (X) = g(S^7)
\ee

We write the round metric on $S^7$ as
\be
g({S^7}) = d\beta^2 + \sin^2 \beta\ g(S^3)+ \cos^2\beta\  g'(S^3)
%d\Omega_3^2 + \cos^2 \beta  d{\Omega'}_3^{2}
\ee
where $g(S^3)$ and $g'(S^3)$
%$d\Omega_3^2$ and $d{\Omega'}_3^{2}$ 
are round metrics on $S^3$ with unit radius
and  $0\le \beta \le{\pi\over 2}$. 
We now divide by the $U(1)$ which acts nontrivially on the
first $S^3$ as the standard fixed point free
Hopf action. In this case
\be
g_6 (Y) = d\beta^2 + {1 \over 4}\sin^2 \beta\ g(S^2) + \cos^2\beta\  g'(S^3)
%+ {1 \over 4}\sin^2 \beta d\Omega_2^2 + \cos^2\beta  d\Omega_3^{`2}
\ee
where  $g(S^2)$
%$d\Omega_2^2$ 
is the round metric on $S^2$ of radius one,
$h(Y) = \sin \beta$ 
and $A$ is the standard charge one Dirac monopole
connection on $S^2$. Note that
$g_6 (Y)$ is a squashed metric on $S^6$ --- the round metric does not have the
factor of $1 \over 4$.

In $S^7$ the $U(1)$ has a fixed point at $\beta = 0$
which is a copy of $S^3$.
Since this is codimension four in $M$  theory, in Type IIA theory
we have a $D$6-brane. Thus, our solution describes $N$ $D$2-branes
on a cone over $S^6$ which has been deformed by the presence of
a $D$6-brane wrapping a cone over $S^3$. In the large $N$ limit
we get IIA on $S^6 \times AdS_4$ with a space filling $D$6-brane
wrapping $S^3 \times AdS_4$. Note that since $S^3$ is contractible in $S^6$,
the total $D$6-brane charge is zero: {\it i.e.} the wrapped $D$6 is dipolar.

Since a cone on $S^6$ is ${\bf R^7}$ we interpret the solution as the result
of adding $N$ $D$2-branes to a $D$6-brane in flat space, with the $D$2-branes
inside the $D$6-brane.
The deviation of the metric on $S^6$ away from the round metric is precisely
the back reaction of the metric in the presence of the $D$6-brane. Note, however,
that this interpretation makes sense only near the branes, because $D$-branes
in flat space should approach flat space at infinity, with a constant
dilaton.

The complete supergravity solution describing $N$ $D$2 branes inside 
a $D$6-brane
was found in \cite{cherkis}. Near the branes it is exactly the
solution described above.

Many of the features of this example will be present in all the
examples we discuss later, so it is worthwhile summarising them.
The first and most important for our purposes is that the IIA duals of
Freund-Rubin compactifications are only good descriptions of configurations
of $D$2-branes and $D$6-branes {\it near} the branes. The above
simple example illustrates that --- since the metric of $D$-branes
in flat spacetime should asymptote to flat spacetime --- the
$M$2-brane metric is not a good description of the Type IIA
background away from the branes. Since our main interest
is in any case Freund-Rubin compactification, we only require
the solution near the branes\ft{As an aside, our results
do predict the existence of many new manifolds of $Spin(7)$ holonomy,
which represent the metric transverse to the $M$2-branes in the
full solution.}.

A second striking feature which will persist in later examples is that, in
Type IIA on $S^6 \times AdS_4$, the $D$6-brane is wrapping a contractible
submanifold, yet fills the entire $AdS$ space. This does not lead to
a tadpole because the $D$6-brane charge is zero. However it raises an obvious
stability issue. Namely, why doesn't the brane simply slip off of the
sphere and decay to the vacuum? This cannot happen either classically
or semiclassically because of supersymmetry. A supersymmetric vacuum state
with negative cosmological constant is classically stable against decay
due to the Breitenlohner-Freedman bound \cite{bf}. Similarly, supersymmetry
prohibits the existence of spherical instantons which mediate 
%bubble of the would-be-true vacuum,  so there is no possibility of tunneling 
semi-classical tunneling processes out of the supersymmetric vacua \cite{cdl,weinberg,cvetic}. 
See \cite{banks}
for some recent discussion of bubble nucleation in quantum gravity.
Solutions describing branes wrapping contractible submanifolds
have appeared in other $AdS$ contexts \cite{karch}.

%\newpage
\bigskip
\Large
\noindent
{\bf {\sf 3. IIA Duals from $G_2$-cones}}
\normalsize
\bigskip

One possible approach to obtaining compactifications with chiral fermions
is to find dual Type IIA descriptions in which  $D$6-branes intersect along $AdS_4$
in such a way that the open string stretched between them is a chiral
fermion. This approach was certainly useful in the case of compactifications
to Minkowski space;  for the first supersymmetric chiral examples 
see \cite{Cvetic:2001nr}.

Since $D$6-branes which intersect along a copy of 
four dimensional Minkow\-ski spacetime 
are known to typically give rise to chiral fermions \cite{bdl},
one's first guess is to generalise the previous example by adding an additional
stack of $D$6-branes intersecting the first. Remarkably the supergravity 
solution
for this system, even including a large number of $M$2-branes, 
is already known in the near horizon limit \cite{bilal}
and we will rederive it here in a simpler way.
Unfortunately, as we will explain, even though this spacetime does contain
chiral fermions, it also contains their $C$-conjugates, so the total
spectrum is non-chiral.

Consider a pair of flat $D$6-branes in flat spacetime,
${\bf C^3}{\times}{\bf R^{3,1}}$ which intersect along the copy of
${\bf R^{3,1}}$ at the origin of ${\bf C^3}$ and are oriented in such a way
that the three angles between them are all ${2\pi \over 3}$. This is a
supersymmetric configuration and the massless open strings stretched between
the two branes are described by a chiral supermultiplet containing a
chiral fermion. Near the intersection point this Type IIA background is
known to be dual to $M$ theory on a $G_2$-holonomy cone over ${\bf CP^3}$
\cite{aw}.
In $M$ theory, the conical singularity of the $G_2$-holonomy metric 
\be
d\rho^2 + \rho^2 g({\bf CP^3})
\ee
corresponds to the intersection point of the two branes\ft{$g({\bf CP^3})$
is not the Fubini-Study metric \cite{brysal}.}.

Since it has a Type IIA interpretation, the metric $g_6$ on ${\bf CP^3}$
can be written as
\be
g ({\bf CP^3}) = f^2 (S^5) (d\tau + A)^2 + g (S^5)
\ee
{\it i.e.} as a circle fibration over $S^5$. 
Note that the induced metric on $S^5$
is not round, due to the presence of the branes. The metric can be
given explicitly, but we will not require its details. 

In $M$ theory, the eleven dimensional metric describing these branes is
\be
g_{10+1} = g_{2+1} + dw^2 +  d\rho^2 + \rho^2 g({\bf CP^3}) 
\ee
with $w$ a coordinate on ${\bf R}$.
Note that this spacetime can be written in the form
\be
g_{10+1} = g_{2+1} + dr^2 + r^2 (d\alpha^2 + \sin^2 \alpha\  
g({\bf CP^3}))
\ee
where $w = r\cos\alpha$ and $\rho = r\sin\alpha$, with $0 \leq \alpha 
\leq \pi$. 

We can now add $N$ $M$2-branes at $r = 0$, giving a solution 
which takes the form of $(1)$.
The near horizon limit of this solution is
\be
g_{10+1} = g(AdS_4 ) + a^2 
\bigl( d\alpha^2 + \sin^2 \alpha\  g ({\bf CP^3})\bigr) 
\ee
This is a Freund-Rubin compactification on the 7-manifold with metric
of the form $g_7 (X) = d\alpha^2 + \sin^2 \alpha g_6 $. 
This Freund-Rubin solution was recently
found in \cite{bilal}. Notice that near $\alpha = 0$ and $\pi$ the metric
has conical singularities isomorphic to those of the original $G_2$-cone
we began with. Since we already know that such  singularities support
chiral fermions in $M$ theory \cite{aw} this Freund-Rubin solution has
two chiral fermions. However, these two chiral fermions are $C$-conjugates
of each other because the two singularities have opposite orientation.
Hence the full spectrum is non-chiral. We now discuss the IIA dual.

Near the branes the Type IIA metric is
\be
g_{IIA} = a^3 f(S^5) \sin\alpha \bigl(a^{-6} r^4 g_{2+1} + r^{-2}dr^2
+d\alpha^2 + \sin^2 \alpha\,  g (S^5)\bigr)
\ee
which describes a compactification to $AdS_4$ on a non-round
$S^6$ with two $D$6-branes. The $M$ theory circle vanishes when
the function $f$ is zero, and this consists of two copies of $S^3$.
Thus the $D$6-branes each wrap different $S^3$'s in $S^6$ and meet each
other at $\alpha = 0$ and $\pi$. 

This is also clear because the space surrounding the $D$2-brane is $S^6$,
with or without $D$6-branes. The two $S^3$'s are simply the intersections
of  ${\bf R^4}$'s in ${\bf R^7}$ with $S^6$. We can thus interpret
the solution as describing the result of adding a large number
of $D$2-branes to the two intersecting $D$6-branes in flat spacetime.
Again we emphasise that the solution is only valid near the $D$6-branes.

Near each intersection point the
geometry is that of the intersecting branes in flat space, so again we
see that there are two ($C$-conjugate) chiral fermions in this
description, consistent with the $M$ theory interpretation. 

Much more generally, if
\be
d\rho^2 + \rho^2 g_6 (W)
\ee
is {\it any} $G_2$-holonomy cone, then
\be
g_7 (W) = d\alpha^2 + \sin^2 \alpha\, g_6 (W)
\ee
is a compact Einstein manifold with positive cosmological constant 
\cite{bilal} which provides a supersymmetric Freund-Rubin solution
of $M$ theory.

Hence, we can write an $M$2-brane solution whose near horizon limit
is a Freund-Rubin compactification with a pair of $C$-conjugate
conical singularities. Many examples of such $M$ theory vacua can be
produced by simply considering Type IIA on flat 
${\bf C^3}{\times}{\bf R^{3,1}}$ with collections of $D$6-branes
spanning linearly embedded ${\bf R^3}$'s through the origin in ${\bf C^3}$.
If the ${\bf R^3}$'s are chosen at $SU(3)$ angles to one another, then near
the intersection point the $M$ theory description is a conical singularity
in a $G_2$-manifold. 

For instance, if $W = SU(3)/{U(1)^2}$ with its natural
metric, then the corresponding $G_2$-holonomy cone is the $M$ theory
description of 3 $D$6-branes at $SU(3)$ angles \cite{aw}. 
If $W = {\bf WCP^3_{ppqq}}$
then the IIA background has $p$ $D$6-branes intersecting $q$ $D$6-branes
again at $SU(3)$ angles \cite{aw}, although in this case the metric on $W$ is not
known.

There is another simple reason why beginning with supersymmetric $D$6-branes
in flat space does not produce chiral fermions in $M$ theory when we
add a large number of $M$2-branes. This is because in the $AdS$ region,
the $D$6-branes are
wrapping 3-manifolds in $S^6$. Such a 3-manifold has trivial homology class
and hence the intersection number of a pair of such 3-manifolds, which 
is what counts the net number of chiral fermions, is zero. Hence, in order to
produce genuinely chiral examples we need to find Type IIA duals
in which the $S^6$ gets replaced with a 6-manifold $Z$ with non-zero
third homology group. This can be achieved by replacing the
${\bf R^7}$ transverse to the $D$2-branes in flat space by a more
complicated manifold. We will in fact replace ${\bf R^7}$ by a $G_2$-holonomy
cone over a 6-manifold $Z$, for which $H_3 (Z)$ is non-trivial.

Note that now there are two $G_2$-holonomy cones playing distinct roles
in these spacetimes. One has a base $W$ and represents the $M$ theory
singularity close to the intersection of the $D$6-branes. The second has
base $Z$ and represents the space transverse to the $D$2-branes and
which the $D$6-branes wrap.

The basic overall picture of the Type IIA backgrounds we will henceforth
consider may be summarised as follows. We begin with Type IIA on
a $G_2$-holonomy cone $C(Z)$ with base $Z$ times 3d Minkowski spacetime.
We then add supersymmetric configurations of $D$2-branes and $D$6-branes.
The $D$2-branes span the Minkowski space and reside at the tip
of the cone over $Z$. Each set of $D$6-branes wrap the Minkowski space
times a supersymmetric 4-manifold $N_i$ in $C(Z)$. We additionally
require that $N_i$ is itself a cone $C(Q_i)$ with $Q_i$ a compact 3-manifold in
$Z$. The complete supergravity solution describing this background is
difficult to write, but it is clear that it should share many features of the
the previous backgrounds we have discussed, in which $Z=S^6$. In particular,
supersymmetry guarantees its existence.

Specifically, the spacetime will asymptote to $C(Z)$ times Minkowski space,
without branes.
Near the tip of the cone we will find a Type IIA compactification on
$Z$ to $AdS_4$. Additionally, there will be $D$6-branes which fill the
$AdS$ space and wrap the 3-manifolds $Q_i$ in $Z$. In order to obtain chiral 
fermions in this background we require the intersection numbers
of the $Q_i$ to be non-zero. 

In $M$ theory the complete spacetime will be asymptotic to
$S^1 \times C(Z)$ times 3d Minkowski space. Without the $M$2-branes
the point at which all the $D$6-branes meet --- the tip of $C(Z)$ ---
is described in $M$ theory by a conical singularity in a manifold
with $Spin(7)$-holonomy. Hence,
near the $M$2-branes
the $M$ theory metric
will approach a Freund-Rubin solution of the form $(2)$, in which
$X$ is a compact 7-manifold with a $U(1)$ quotient which is $Z$.
The $U(1)$ fibers will degenerate to zero size on each of the
$Q_i$'s. At the intersection points between $Q_i$ and $Q_j$ $X$
will have a conical singularity with base $W_{ij}$. Near the
intersection points the cone on $W_{ij}$ also has $G_2$-holonomy.
More generally there could be multiple intersection points.

Supersymmetry requires that the complete 8-manifold $V$ transverse
to the $M$2-branes is a manifold with $Spin(7)$-holonomy. We are thus
predicting the existence of many new metrics of $Spin(7)$-holonomy.
$Spin(7)$-manifolds interpolating between $S^1 \times C(Z)$ and
cones on Freund-Rubin manifolds $X$ are known \cite{gibb,serg} for
some choices of $Z$ and $X$, though not in the cases we will study here.

An important new feature of these more general backgrounds is that
the $Q_i$ are incontractible, so we cannot wrap $D$6-branes on arbitrary
$Q_i$, since $Z$ is compact. Cancellation of tadpoles requires that
the total $D$6-brane charge is zero, {\it i.e.}
\be
\sum_i \;\; k_i [Q_i] = 0
\ee
where $k_i$ is the number of $D$6-branes wrapping $Q_i$ whose homology class
is $[Q_i]$. In compactifications to flat space, for example on a Calabi-Yau,
this condition can never be satisfied non-trivially
in a supersymmetric fashion because there
the $Q_i$ are calibrated by a closed 3-form. The way around this problem
there is to introduce orientifold planes which have negative charge.
As we will see in compactifications to $AdS_4$ on $Z$ one can find non-trivial
solutions to $(19)$ {\it without} orientifolds, and in this case $Q$ is 
{\it not} calibrated by a closed 3-form. 
These examples are presumably related to the 
``generalised'' calibrations studied in \cite{gencal}.

Like the solution in section two, 
these configurations, since they have zero $D$6-brane
charge are in principle non-perturbatively unstable to decay to an $AdS$
vacuum without $D$6-branes, but as discussed there, these vacua are, 
at the very least,  semi-classically stable.

The only known example of a $G_2$-holonomy cone $C(Z)$ in which
$Z$ has non-trivial 3-cycles is $S^3 \times S^3$ (or its discrete quotients).
Happily, as we will see, 
supersymmetric compactifications of Type IIA on $S^3 \times S^3$
with $D$6-branes exist which contain chiral fermions.

%Fortunately we will be able to show that supersymmetric $D$6-brane 
%configurations
%exist with chiral fermions.

%\newpage
\bigskip
\Large
\noindent
{\bf {\sf 4. Chiral Fermions in $S^3 \times S^3 \times AdS_4$}}
\normalsize
\bigskip

The following metric has $G_2$-holonomy \cite{brysal}
\be
ds^2 = dr^2 + {r^2 \over 9} ( \omega_a^2 + {\tilde \omega_a}^2 
-\omega_a {\tilde \omega}_a )
\ee
This is a cone with base $Z = S^3 \times S^3 $, where the
induced metric on $Z$ is $G/H$, with $G =  SU(2)^3$ and $H$ the
diagonal $SU(2)$ subgroup. The
$\omega_a$ are left invariant 1-forms on the first $S^3$, viewed
as the  $SU(2)$ group manifold, and
${\tilde \omega_a}$ are left invariant 1-forms on the second $S^3$.
The metric is invariant under the order six
group generated by exchanging the two $S^3$'s which acts
on the one forms as
\be
\alpha: \;\; 
%\omega_a \leftrightarrow {\tilde \omega}_a
\left(\matrix{\omega_a\cr  {\tilde \omega}_a }\right) 
\rightarrow 
%\left(\matrix{0&1\cr1&0\cr}\right)\left(\matrix{\omega_a\cr  {\tilde \omega}_a }\right)
\left(\matrix{ {\tilde \omega}_a \cr \omega_a}\right) 
\ee
and by cyclic permutations of the three $SU(2)$'s in $G$, which has the effect
\be
\beta: \;\; \left(\matrix{\omega_a\cr  {\tilde \omega}_a }\right) 
\rightarrow 
%\left(\begin{array}{cc} 0 & -1\\1 & -1\end{array}\right)
\left(\matrix{-{\tilde \omega}_a\cr \omega_a- {\tilde \omega}_a}\right) 
\ee

The metric on $S^3 \times S^3$ has an almost complex
structure defined by introducing the complex frames
\be
\eta_a \equiv {1 \over 3}(\omega_a + \tau{\tilde \omega_a})
\ee
where $\tau = e^{2\pi i \over 3}$. The metric is thus
\be
{1 \over 9} ( \omega_a^2 + {\tilde \omega_a}^2 
-\omega_a {\tilde \omega}_a ) = \eta_a {\bar \eta}_a
\ee
The associated Kahler and holomorphic volume form are
defined as
\ba
\omega &\equiv& {i \over 2} \eta_a \wedge {\bar \eta_a}\\
\Omega &\equiv& \eta_1\wedge \eta_2 \wedge \eta_3
\ea
and obey the equations
\ba
d \omega &=& -3{\rm Im}\;\Omega \\
d\mbox{Re}\;\Omega &=& -2 \omega\wedge \omega
\ea

On the $G_2$-holonomy cone $dr^2 + r^2 \eta_a {\bar \eta_a}$ 
there exists a covariantly constant 3-form $\varphi$ and
4-form $*\varphi$. These forms may be written in terms of 
$\omega$ and $\Omega$ as follows
\be
\varphi =  r^2 dr \wedge \omega -r^3 {\rm Im}\;\Omega
\ee
and
\be
*\varphi = r^3 {\rm Re}\;\Omega \wedge dr +{r^4 \over 2}\omega \wedge \omega
\ee

Note that in this basis
\be
\alpha: \eta_a \rightarrow \tau {\bar \eta_a}
\ee
and $\beta$ acts holomorphically as multiplication by
$\tau$. Clearly, extending these discrete symmetries 
trivially to the cone, 
\be
\alpha: \varphi \rightarrow - \varphi
\ee
leaving $*\varphi$ invariant, whereas $\beta$ preserves both
$\varphi$ and $*\varphi$.

We are interested in four dimensional, supersymmetric submanifolds $N$ of the
$G_2$-holonomy cone. These are defined by the condition
\be
*\varphi|_N = \pm \mbox{dVol}(N)
\ee
{\it i.e.} that the restriction of $*\varphi$ to $N$ is its volume
form, up to orientation. If $N_1$ and $N_2$ are calibrated with opposite
orientation, then a configuration of branes wrapping $N_1$ and anti-branes
wrapping $N_2$ is supersymmetric. Equivalently, a configuration of branes
on $N_1$ and branes on ${\bar N_2}$ is supersymmetric, where ${\bar N_2}$
is $N_2$ with the opposite orientation. Note that, up to the overall orientation of $N$, $(33)$ 
is equivalent to
\be
\varphi|_N = 0
\ee

Examples of supersymmetric $N$'s may be obtained as the fixed
point sets of orientation reversing isometries which preserve
$*\varphi$ but reverse the sign of $\varphi$. Clearly
$\alpha$ is just such an isometry. In $S^3 \times S^3$, the
fixed point set of $\alpha$ is the `diagonal' $S^3$. Therefore
in the cone on $S^3 \times S^3$ the $\alpha$ fixed points
are a cone over this $S^3$, which is a copy of ${\bf R^4}$.
The induced metric on this ${\bf R^4}$ is not flat.

From the formula for $*\varphi$, one can explicitly check
that the restriction to this ${\bf R^4}$ is the
volume form, up to orientation. Thus, after taking the
large $N$ limit, we learn that in Type IIA
on $Z \times AdS_4$, the diagonal $S^3$ is supersymmetric.

Furthermore, since $\varphi$ and $*\varphi$ are $\beta$ invariant, 
the images of the diagonal $S^3$ under $\beta$ and
$\beta^2$ are also supersymmetric. So we have found three
supersymmetric $S^3$'s  in Type IIA on $S^3 \times S^3 \times AdS_4$.
Moreover, the union of all three of these $S^3$'s is supersymmetric.
Notice that the $S^3$'s are indeed `calibrated' by ${\rm Re}\;\Omega$ which is
not closed and that they each come in a three dimensional family parametrised
by $SU(2)$. 
Denote these three families of supersymmetric $S^3$'s by $Q_{\alpha}$,
$Q_{\beta\alpha}$ and $Q_{\beta^2 \alpha}$.

Before we go on to discuss wrapped $D$6-branes we need to consider
the third homology of $ Z = S^3 \times S^3$, 
$H_3(Z, {\bf Z}) \cong {\bf Z \oplus Z}$. Hence the third homology
is a rank two lattice. The homology can be generated by $(1,0)$ and
$(0,1)$ and explicit representatives of the generators may simply be taken
to be $S^3 \times {pt}$ and ${pt} \times S^3$. Since $\alpha$ and $\beta$
are symmetries of $Z$, they must also act naturally on $H_3(Z, {\bf Z})$.
This action is readily seen to be represented by the matrices
\be
\alpha^* = \left(\begin{array}{cc} 0 & 1\\1 & 0\end{array}\right)
\ee
and
\be
\beta^* = \left(\begin{array}{cc} 0 & -1\\1 & -1\end{array}\right)
\ee
Because of these symmetries $H_3(Z, {\bf Z})$ may naturally be regarded
as the root lattice of the $SU(3)$ Lie algebra, {\it i.e.} the metric on $Z$ induces
the lattice with a natural metric and complex structure. When regarded
as a lattice in ${\bf C}$, the metric is
\be
%ds^2 = 
dzd{\bar z}
\ee
where $z = x + \tau y$. Then $\beta$ acts by ${2\pi \over 3}$
rotations and $\alpha$ by complex conjugation.

The fixed point set of $\beta$ acting on $H_3(Z, {\bf Z})$ is empty. Therefore
if we take any 3-cycle and its two images under $\beta$, their sum is zero in
homology.
This means that 
the homology classes of the three mutually supersymmetric $S^3$'s
add up to zero. Hence they are either given  by $[Q_{\alpha}] = (-1,-1)$,
$[Q_{\beta\alpha}] = (1,0)$, 
$[Q_{\beta^2 \alpha}] = (0,1)$ {\it or} by $[Q_{\alpha}] = (1,1)$,
$[Q_{\beta\alpha}] = (-1,0)$, 
$[Q_{\beta^2 \alpha}] = (0,-1)$. There are thus two different
collections of mutually supersymmetric $S^3$'s. The difference
between these cases can be described physically as the difference
between branes and anti-branes.

Finally we discuss the intersection form between 3-cycles.
If $[Q_1]$ =  $(a, b)$ and $[Q_2]$ = $(c, d)$ are two three-cycles, 
then their intersection 
number  is
\be
[Q_1 ]\cdot [Q_2 ] = - [Q_2 ]\cdot [Q_1 ] 
= \det\left(\begin{array}{cc} a & b\\c & d\end{array}\right) 
\ee
Note that the pairwise intersection numbers between the
three supersymmetric $S^3$'s is $\pm 1$. 

Since the homology classes of the three supersymmetric cycles add up
to zero, the tadpole cancellation condition may be satisfied
by wrapping an equal number $k$ of $D$6-branes on each 3-cycle.

There is a simple model of
these $D$6-branes in flat space which is a good local description of
the brane configuration.
For this, replace $Z = S^3 \times S^3$ with ${\bf C^3}$ and
with flat metric
\be
%g_6 = 
dz_a d{\bar z}_a
\ee
where the complex structure is defined by $z_a = x_a + \tau y_a$.
We first add a $D$6-brane along the 3-plane at $y_a = 0$.
This is the local analog of the $(1, 0)$ brane on $Z$. Now we
act with the center of $SU(3)$ on this 3-plane and put a $D$6-brane
at the images. The center of $SU(3)$ is generated by $\tau{\bf 1}$.
This gives a configuration of 3 $D$6-branes which are supersymmetric
because they are related by $SU(3)$ rotations. The $\tau$ and
$\tau^2$ images  are the analogs of the images of $(1, 0)$ under
$\beta$. The image under $\tau$ is the analog of $(0, 1)$
and the image under $\tau^2$ is the analog of $(-1,-1)$. The minus sign
is because $SU(3)$ preserves orientation. Notice that the
$\tau^2$ brane may equally well be regarded as an {\it anti-brane}
at 60 degrees to the other two. 

In fact one may check that
the three supersymmetric $S^3$'s in $Z$ meet at precisely $120$ degrees
in all directions, so the flat space model is indeed a good description of
the branes locally.

Near the intersection point of the three sets of branes
the $M$ theory description is that of a conical singularity
with base $W$. In this case $W$ is a ${\bf Z_k}$-orbifold of $SU(3)/{U(1)^2}$
\cite{aw}.

The spectrum of light charged particles in this model is simple to deduce.
The gauge group is $U(k)^3$ and there are three sets of bifundamental
fields in the representation ${\bf (k,{\bar k}, 1) \oplus (1,k, {\bar k})
\oplus ({\bar k}, 1, k)}$. The reader might be worried that we have extrapolated
results for the spectrum based upon those for  which the branes intersect
in flat spacetime. However, at extremely large $N$ the IIA spacetime is
almost flat and the net number of chiral fermions is independent of $N$.

Apart from the central $U(1)^3$, this gauge group and matter content is anomaly free. 
This is expected because
the brane configuration is free from tadpoles, at least in the supergravity 
approximation. As far as the anomolous $U(1)$'s are concerned we expect that
they are in fact massive in the quantum theory, as is the case in all known
string compactifications with anomolous $U(1)$'s at tree level \cite{anom}.

There may be many other supersymmetric configurations of $D$6-branes
on $Z$. For example, let
$(p_i, q_i )$ be three 3-cycles. We may then wrap $k_i$ $D$6-branes
on $(p_i, q_i)$ if 
\be
\sum_i k_i (p_i, q_i ) = (0, 0)
\ee
This is readily seen to give an anomaly free spectrum of chiral fermions,
charged under the gauge group $SU(k_1) \times SU(k_2) \times SU(k_3)$.
The matter spectrum here is given by 
$n_{12} {\bf (k_1, {\bar k_2}, 1)} \oplus 
n_{23} {\bf (1,{k_2}, {\bar k_3})} \oplus n_{31}{\bf ({\bar k_1},1, {k_3})}$,
where
\be
n_{ij} = p_i q_j - p_j q_i
\ee
Unfortunately, 
we have not been able to construct supersymmetric 3-cycles whose classes
are not multiples of $(1, 0)$ and its images under $\beta$. However
if they existed one might be able to find more phenomenologically interesting
supersymmetric models in this way.
We think it is also likely that
many other Freund-Rubin compactifiations exist which give rise
to more phenomenologically interesting models of particle physics.
These may or may not be dual to Type IIA vacua.

\bigskip
%\Large
\noindent
{\bf {\sf 4.1. Compactifications?}}
\normalsize
\bigskip

An important point concerning typical Freund-Rubin 
compactifications
such as those on round spheres or other homogeneous spaces is that there
is no consistent low energy limit in which one can ignore the Kaluza-Klein
modes. The reason is simply that in such examples the scale  of $X$
is of the same order as that of the $AdS$ spacetime and hence fluctuations of
the supergravity multiplet have masses of the same order as
the Kaluza-Klein modes. This problem can be avoided if the volume
of $X$ computed in the metric $g_7(X)$ in equation $(2)$ is parametrically
smaller than that of the round $S^7$ of unit radius, whilst keeping
the cosmological constant equal  to $6$.
Note that
examples of Freund-Rubin 7-manifolds obeying this criterium, but
preserving ${\cal N}=2$ spacetime supersymmetry,
may be found in \cite{herzog}, and we presume that they also exist in the
${\cal N}=1$ cases of interest here. 

In the examples dual to Type IIA vacua with branes that we presented in this 
section this criterium is probably not obeyed,  although, since we
have not obtained the full supergravity solution in these examples
we cannot be sure.

One method for reducing the volume of $X$ whilst preserving its cosmological 
constant is to consider discrete quotients of $X$, since if $X/\Gamma$ is
a discrete quotient of $X$, then ${\rm Vol}({X/\Gamma})$ = ${\rm Vol}(X)/
|\Gamma|$, where $|\Gamma |$ is the order of $\Gamma$. Note however
that we require $\Gamma$ to be such that $X/\Gamma$ is smaller
than $X$ in {\it all} directions in order to increase the mass of all
Kaluza-Klein modes. In the examples presented in this section,
$Z$ has $SU(2)^3$ symmetry group and we can take $\Gamma$ to be
any discrete subgroup. The largest freely acting subgroup that
$Z$ has, which reduces the volume equally in all
directions,  is the product of two copies of the
binary icosahedral group ${\cal I}$, which acts in such a way that
$Z/\Gamma = 
{S^3}/{\cal I}{\times}{S^3}/{\cal I}$. Since $|{\cal I}| = 120$
the volume of the Type IIA spacetime is reduced by a factor of $120^2$,
thus reducing the typical radius by a factor of about 5.
This accounts for six of the seven dimensions of $X$. The last dimension
is the $U(1)$ fiber which vanishes on the $D$6-branes. Since we have a $U(1)$
symmetry we can gauge a discrete subgroup of order $M$ and this reduces
the radius of this dimension by $M$. Note that this increases the
number of $D$6-branes by a factor of $M$.
A factor of $5$ is just about enough to give a gap in the spectrum between
the Kaluza-Klein modes and the supergravity, gauge and matter multiplets.

In this case one obtains supersymmetric, chiral Type IIA compactifications
with the branes wrapping copies of ${S^3}/{\cal I}
$. One then has interesting
additional possibilities for breaking the $SU(k)$ gauge groups
by discrete Wilson lines, which might be worthy of further study.

\bigskip
\Large
\noindent
{\bf {\sf 5. Holographic Duals}}
\normalsize
\bigskip

One of the most interesting aspects of Freund-Rubin compactifications
is that they are dual to three dimensional field theories.
For example, if $X$ is such that its singularities support a copy of
the supersymmetric standard model, then there exists a 3d field theory on
the worldvolume of the $N$ $M$2-branes which is the holographic dual
of $M$ theory on $X \times AdS_4$. It is clearly of importance then
to understand in detail these three dimensional field theories.
In the examples of section three
we can actually say what this 3d theory is in the UV.

Those examples have dual Type IIA descriptions as $D$2-branes in the
background of intersecting $D$6-branes in flat spacetime, and in those
cases the 3d theory is simply the worldvolume theory of the $D$2-branes.
In fact, in these cases, the 3d theory on the $D$2-branes has already
been studied in \cite{dave}. Since we have found
$AdS$ duals of these theories at large $N$, all of these theories
must flow in the infrared to renormalisation group fixed points.
An interesting point for further study is that
in some of these examples, the gauge dynamics in the bulk of
$AdS_4$ can be strong at low energies. For example
in the case that we begin with a set of $p$ $D$6-branes
intersecting a single $D$6-brane at ${2\pi \over 3}$ angles,
the gauge group and matter content supported at the singularities
of $X$ is $SU(p)$ with one flavour. In flat space (which here
is the large $N$ limit), this gauge theory is known to generate
a non-perturbative superpotential \cite{affleck}. 
If there is no mass term for the
quarks and their superpartners this gauge theory has no vacuum.
This model already raises many questions when embedded into a Freund-Rubin
compactification of $M$ theory: is a mass term generated by quantum
effects such as membrane instantons? Can the non-perturbative superpotential
be understood in terms of membrane instantons? Since the background
has a classical cosmological constant the superpotential is classically
non-zero,
so what is the vacuum state in $M$ theory?

In the examples of section four which contain chiral fermions in
$AdS_4$, it is more difficult to understand the $D$2-brane theory.
However, in principle one can quantise open strings in the $G_2$-holonomy
background perturbatively and hence understand the $D$2-brane field
theory in the UV. There are some obvious properties that this theory has.

The gauge group is $U(N)$. 
Since the three sets of $k$ $D$6-branes each support
$SU(k)$ gauge groups, the d=3 theory has a $G=SU(k)^3$ {\it global} symmetry.
Furthermore, the strings stretched between the $D$2-branes and $D$6-branes
are massless in the UV and give rise to three sets of $k$ component
fundamentals of $U(N)$, transforming in the obvious way under $G$. The
UV theory is supersymmetric, so all these fields are in matter supermultiplets
of d=3 ${\cal N}$ $=1$ supersymmetry.

Holography might also provide an interesting way to
look at supersymmetry breaking. For, instance, if $X$ has singularities 
supporting a gauge theory which is known to spontaneously break supersymmetry,
then how is that manifested in the holographically dual field theory?
The UV/IR correspondence suggests that if supersymmetry is restored
at high energies in the 4d bulk (at extremely large $N$), then the
dual field theory has supersymmetry in the IR limit, but is broken in
the UV. For example, the dual theory 
might be a supersymmetric theory perturbed in the UV by
non-supersymmetric irrelevant operators.

Finally, if the quantum dynamics of $M$ theory on $X$ is such
that, after supersymmetry breaking is correctly accounted for, the
vacuum is de Sitter space, we might be fortunate enough to learn something
about the holographic duals of inflationary cosmologies. One possible context
where this idea might be realised is if the singularities of $X$ produce
strong gauge dynamics which spontaneously break supersymmetry at a scale
$m_{susy}$. If $m^4_{susy} \geq -\Lambda_{AdS} \sim m_p^2 a^{-2}$ 
then the effective cosmological
constant will be $\geq 0$. For example, if $m_{susy} \sim 10^{11}$GeV
then the original $AdS$ mass scale $a^{-1}$ is of order 1 TeV.

\bigskip
\Large
\noindent
{\bf {\sf Acknowledgements}}
\normalsize
\bigskip

We would like to thank Tom Banks, Mike Douglas, Dan Kabat,
Juan Maldacena and Greg Moore
for useful discussions. We
would also like to thank Dan Kabat and 
the Department of Physics at Columbia
University for hospitality where this work was completed.

\end{document}